\documentclass[12pt,english,aps,manuscript]{article}
\usepackage[T1]{fontenc}
\usepackage[latin1]{inputenc}
\usepackage{geometry}
\usepackage[active]{srcltx}
\usepackage{amsmath}
\usepackage{amssymb}
\usepackage{graphicx}
\usepackage[numbers]{natbib}

\makeatletter

\usepackage{geometry}

\makeatletter

\usepackage{color}

\makeatletter
\newcommand{\bee}{\begin{equation}}
\newcommand{\ee}{\end{equation}}
\newcommand{\beea}{\begin{eqnarray}}
\newcommand{\eea}{\end{eqnarray}}

\makeatother

\makeatother

\makeatother

\usepackage{babel}
\begin{document}
\begin{center}
\textbf{\Large One PI and Wilsonian Actions in SUSY theories }
\par\end{center}{\Large \par}

\begin{center}
\vspace{0.3cm}
 
\par\end{center}

\begin{center}
{\large S. P. de Alwis$^{\dagger}$ }
\par\end{center}{\large \par}

\begin{center}
Physics Department, University of Colorado, \\
 Boulder, CO 80309 USA 
\par\end{center}

\begin{center}
\vspace{0.3cm}
 
\par\end{center}

\begin{center}
\textbf{Abstract} 
\par\end{center}

The soft breaking terms in supersymmetric theories are calculated
at some high scale characterizing the hidden supersymmetry breaking
sector, and then evolved down to the TeV scale. These parameters are
usually presented as the ones that should be compared to experiment.
The physical parameters however are those occurring in the quantum
effective (1PI) action - in particular the physical mass is the location
of the pole in the full quantum propagator. Here we discuss the relation
between the two and the possible existence of additive contributions
to the gaugino mass. We argue that infra red effects which violate
non-renormalization theorems are absent (for the 1PI action) if the
calculation is done at a generic point in field space so that an effective
IR cutoff is present. It follows that if a gaugino mass term is absent
in the Wilsonian action it is absent in the 1PI action. 

\begin{center}
\vspace{0.3cm}
 
\par\end{center}

\vfill{}

$^{\dagger}$ dealwiss@colorado.edu

\eject

\section{Introduction}

In supersymmetric theories the source of supersymmetry breaking is
identified with some hidden sector which is (typically) gravitationally
coupled to the visible sector (i.e. some supersymmetric extension
of the standard model). In such models the scale of the hidden sector
is often close to the Planck scale (gauge mediation being an exception).
In models derived from string theory this scale is identified with
the cutoff scale at which string (or higher dimensional) physics becomes
relevant. The hidden sector theory then determines a set of initial
values for the soft breaking terms of the visible sector.

However these values are clearly not the relevant ones for comparison
with experiment if supersymmetry is responsible for stabilizing the
hierarchy between the weak scale and the Planck (or string or Kaluza-Klein
(KK)) scale. The values of the soft parameters (typically determined
in terms of the gravitino mass $m_{3/2})$, are used as initial values
to solve the supersymmetric RG equations. This gives the values of
these parameters at the TeV scale. These are then to be compared with
the expected experimental values. Indeed it is only after RG evolution
that the Higgs potential develops an instability resulting in the
Higgs mechanism, since the initial values of the Higgs mass squared
parameters are positive.

Nevertheless one may ask how accurate is this identification of the
physical soft masses with what are essentially the running masses
evaluated at the RG scale $\mu\simeq1TeV$. The physical mass of say
the stop is the position of the pole in the full quantum propagator
of the stop. This must take into account not only the effects of high
mass scales but also the effect of functionally integrating over all
low energy scales as well. Nevertheless if indeed there is a physical
stop at the TeV scale, then the difference between this pole mass
and the RG evolved running mass, is expected to be small.

In this paper we will review this in some detail, and then discuss
whether these expectations on the relation between the physics of
the 1PI action and that coming from the Wilsonian action may be violated.
One issue is whether the non-renormalization theorem of SUSY is violated
in the 1PI action when there are massless particles in the theory.
The other is the question of Weyl anomalies and naturalness arguments
for fermion mass renormalization, and whether there could possibly
be an additive contribution to the physical gaugino mass that is not
present in the Wilsonian effective action. We show that this is not
so in either of these cases.

\subsection{Soft mass calculations in SUSY/SUGRA/String theories}

The SUSY breaking which occurs in the hidden sector of supersymmetric
effective field theories is transmitted to the visible sector, usually
taken to be the minimally supersymmetric standard model (MSSM), by
the (gravitational) couplings of the moduli to it. The formulae for
the effective supersymmetry breaking soft term corrections to the
MSSM were worked out in papers going back to the early eighties culminating
in the papers \citep{Kaplunovsky:1993rd,Brignole:1997dp}. 

The {}``classical'' formulae given in these works for the soft terms
are expected to be valid at some high scale just below where the effective
locally supersymmetric field theory gets replaced by a UV complete
theory such as string theory. The structure of the effective theory
is completely determined once the Kähler potential ($K$ a real scalar
superfield), the superpotential ($W$a chiral scalar superfield),
and the gauge coupling function ($f$ a chiral superfield), are given
in terms of the fundamental (super) fields of the theory. In principle
of course if the exact expressions for these, incorporating the effect
of integrating out the quantum fluctuations at scales above the TeV
scale and up to the effective UV cutoff are known, we could use the
formulae of \citep{Kaplunovsky:1993rd,Brignole:1997dp} to compute
the soft masses. 

In practice of course what we do know (in various string theory based
models for instance), are classical expressions corrected usually
by some stringy ($\alpha'$) effects and string loop effects in $K$,
as well as non-perturbative effects in the superpotential. In addition
there are Weyl anomaly (a one-loop effect) contributions to $f$.
Getting to this structure involved integrating out both string states
(with mass scale $M_{s}=\sqrt{1/2\pi\alpha'}$) and Kaluza - Klein
states ($M_{KK}\sim M_{s}/{\cal V}^{1/6}<M_{s}$ with ${\cal V}$
being the volume%
\footnote{Note that ${\cal V}\gg1$ for the effective 4D field theory derived
from string theory via the low energy 10 D SUSGRA to be valid.%
} of the compactified space in string units). Thus we expect these
values for $K,W,f$ and hence the corresponding expressions for the
soft SUSY breaking terms obtained from \citep{Kaplunovsky:1993rd,Brignole:1997dp}
to be valid at some scale just below this KK scale.

In terms of the Planck scale $M_{P}$, the KK scale is given by $M_{KK}\sim g_{s}^{1/2}M_{P}/{\cal V}^{2/3}$,
and depends on the string coupling $g_{s}$ as well as the volume
of compact space, both of which are fixed in terms of internal fluxes,
and choice of the gauge group(s) giving rise to non-perturbative effects.
It is natural however to take this scale to be around the GUT scale
(though it is extremely hard to actually produce an explicit working
GUT model!). In any case even outside of string theory, in SUGRA based
theories of SUSY breaking, the {}``classical'' values of the soft
masses coming from the original data of the effective SUGRA, are taken
to be valid at some high scale close to the GUT scale. The effective
Wilsonian action at the TeV scale is then obtained by using the {}``classical''
data as initial values for solving the RG equations which take care
of the quantum effects in running down to the latter scale. These
are the soft masses that are compared to experiment.

However this procedure does not quite give the physical mass of a
particle. This is defined as the position of the pole in the exact
quantum propagator for the corresponding particle/field ($\phi$ say
with running mass $M(\mu)$). i.e. it is the solution $p^{2}=-M_{phy}^{2}$
of 
\begin{equation}
\Delta^{-1}(p)=p^{2}+M^{2}(\mu)+M^{2}(\mu)\tilde{\Sigma}(-\frac{p^{2}}{\mu^{2}},g_{i}(\mu),\frac{M_{i}^{2}(\mu)}{\mu^{2}})=0.\label{eq:Delta-1}
\end{equation}
 where $\Sigma\equiv M^{2}\tilde{\Sigma}$ is the sum of 1PI diagrams
with two $\phi$ external lines and $g_{i}=g_{i}(\mu),\, M_{i}=M_{i}(\mu)$
stand for the set of running couplings (dimensionless) and masses
evaluated at the scale $\mu$. In writing this we have taken the input
action to be the Wilsonian action $S_{W}(\phi_{i,})$ and the 1PI
action is computed starting from this Wilsonian action. 

In standard renormalization theory, the value of $M_{phys}$ is an
experimental input, however here we are hoping to make predictions
for a set of yet unobserved particles. What the Wilsonian procedure
discussed earlier gives are values for the running masses and couplings
evaluated at the TeV scale, which by assumption are supposed to be
close to the values of the corresponding physical masses. If that
is indeed the case, the former should be a good approximation to the
latter and this is the implicit assumption of typical soft mass calculations
in broken supersymmetry%
\footnote{See for example \citep{Baer:2006rs}.%
}. So in contrast to the procedure in standard renormalization theory,
what is done is to compute the physical masses starting from the input
masses and couplings given in $S_{W}$ valid at the scale $\mu$.
Also the parameters of the Wilsonian action which determine $M_{phy}^{2}$,
are fixed in terms of the initial values $g_{i}^{(0)},M_{i}^{(0)}$
which in turn are determined in terms of the gravitino mass and the
data of the compactification, $\{\sigma_{r}\}$ (Hodge numbers, flux
integers) see equation \eqref{eq:Delta-1}. Hence the physical mass
is given by 
\[
M_{phy}=m_{3/2}(\sigma_{r})h(\sigma_{r})
\]
 (with $h$ a model dependent function of the string/SUGRA data) and
is of course an RG invariant. In practice however what is usually
calculated is $M(\mu)$. The difference may be read off from \eqref{eq:Delta-1};
\begin{equation}
\Delta M^{2}\equiv M_{phy}^{2}-M^{2}(\mu)=M^{2}(\mu)\tilde{\Sigma}(\frac{M_{phy}^{2}}{\mu^{2}},g_{i}(\mu),\frac{M_{i}^{2}(\mu)}{\mu^{2}}),\label{eq:DeltaM}
\end{equation}
and is finite and is at least of one-loop ($g^{2}/4\pi$) order. If
indeed there are physical superpartner particles at the TeV scale
then choosing the RG scale $\mu$ at that scale will minimize the
logarithms that enter into the calculation. 

Now the RHS of (\ref{eq:DeltaM}) does not necessarily imply multiplicative
renormalization - in fact a term of the form $\tilde{\Sigma}\sim\frac{g^{2}}{16\pi^{2}}\frac{\mu^{2}}{M^{2}}$
would give an additive contribution to a scalar mass. In the case
of interest namely when SUSY is softly broken we have an additive
contribution which up to $O(1)$ factors is (assuming there is no
D-term SUSY breaking for simplicity), 
\begin{equation}
\Delta M^{2}\sim\frac{g^{2}}{16\pi^{2}}{\rm sTr}{\cal M}^{2}=\frac{g^{2}}{16\pi^{2}}[2(n-1)m_{3/2}^{2}-2F^{i}\bar{F}^{\bar{j}}R_{i\bar{j}}].\label{eq:DeltaMKL}
\end{equation}
Here $R_{i\bar{j}}=K^{I\bar{J}}R_{i\bar{j}I\bar{J}}$ where the second
factor is the Riemann tensor of the manifold of chiral scalars with
$i\bar{j}$ being tangent to the hidden sector directions (i.e. the
SUSY breaking sector) and $I\bar{J}=1,\ldots,n)$ being the observable
sector directions. The value of the RHS of this equation and hence
the accuracy of the Wilsonian mass is then a model dependent question.
\begin{enumerate}
\item In generic SUGRA hidden sector theories such as mSUGRA we have $\Delta M^{2}\sim\frac{g^{2}}{16\pi^{2}}m_{3/2}^{2}$.
Since the classical soft mass at the UV scale is $O(m_{3/2})$ this
means that the correction in going from the Wilsonian to the physical
mass is formally of the same order as the RG evolution in the Wilsonian
mass. 
\item In extended no-scale models (such as LVS models \citep{Balasubramanian:2005zx},
\citep{Conlon:2005ki}), there is a cancellation between the two terms
on the RHS of \eqref{eq:DeltaMKL} so that $\Delta M^{2}\sim\frac{g^{2}}{16\pi^{2}}\frac{m_{3/2}^{2}}{{\cal V}}$
. Nevertheless the point is that this suppression comes from the suppression
of soft masses in these models so that $M_{cl}^{2}\sim m_{3/2}^{2}/{\cal V}$,
so the contribution due to running (i.e. $M^{2}(\mu)-M_{cl}^{2}$)
is also proportional to $\frac{g^{2}}{16\pi^{2}}\frac{m_{3/2}^{2}}{{\cal V}}$.
\item The third case is the sequestered one in which the gauginos get a
mass in the UV because of the Weyl anomaly and is also therefore a
one-loop effect. But the largest contribution to the scalar masses
(as well as the scalar coupling $A$ term) come from RG running. In
this case $M^{2}(\mu)\sim O(\frac{g^{2}}{16\pi^{2}}m_{3/2}^{2})$,
$ $$\Delta M^{2}\sim\left(\frac{g^{2}}{16\pi^{2}}\right)^{2}m_{3/2}^{2}$
.
\end{enumerate}
Actually though $M^{2}(\mu)-M_{cl}^{2}$ is formally of the same order
as $\Delta M^{2}$ in so far as the leading terms are both one-loop
effects, the former is the result of RG running over many decades
(i.e. integrating over the leading log contributions), and hence the
numerical coefficients (what replaces the large logs in a naive perturbative
contribution) dominate over a one-loop contribution with a small factor
coming from the low energy loop integral in $\Sigma$ i.e. $\sim\ln(M(\mu)/\mu)$.
Thus in all three cases the effect of the RG evolution will be larger
than the loop corrections that go into $\Delta M^{2}$ in \eqref{eq:DeltaM}.

For fermionic masses of course there are naturalness arguments which
imply that $M_{phys}$ will be proportional to $M$. These would apply
if there are no terms (apart from the mass term) that break chirality.
However there are several effects which might affect these arguments
in SUSY theories with massless states. One is the issue of the breakdown
of non-renormalization theories in SUSY theories with massless particles.
The other is that of Weyl anomalies. We discuss these in the next
two sections.

\section{IR effects and Chiral loops }

It has been claimed that the quantum effective action of global supersymmetric
theories with massless particles violates the non-renormalization
theorem for the superpotential \citep{West:1990rm,Jack:1990pd}. Let
us revisit this issue since it is of relevance for the question under
discussion.

The SUSY action for the Wess-Zumino (WZ) model is,
\[
S=\int d^{8}zK(\Phi,\bar{\Phi},V)+\left(\int d^{6}zW(\Phi)+h.c.\right).
\]
The WZ propagators \citep{Buchbinder:1998qv} for chiral scalars are:
\begin{eqnarray}
G_{--}(z,z') & =<\bar{\Phi}(z),\bar{\Phi}(z')>= & \frac{1}{16}D^{2}D'^{2}G_{V}(z,z'),\label{eq:--}\\
G_{+-}(z,z') & =<\Phi(z),\bar{\Phi}(z')>= & \frac{1}{16}\bar{D}^{2}D'^{2}G_{V}(z,z'),\label{eq:+-}\\
G_{-+}(z,z') & =<\bar{\Phi}(z),\Phi(z')>= & \frac{1}{16}D^{2}\bar{D}'^{2}G_{V}(z,z'),\label{eq:-+}\\
G_{++}(z,z') & =<\bar{\Phi}(z),\bar{\Phi}(z')>= & \frac{1}{16}\bar{D}^{2}\bar{D}'^{2}G_{V}(z,z').\label{eq:++}
\end{eqnarray}

$G_{V}$ is the solution (with Feynman boundary conditions) of 
\begin{equation}
(\square-\frac{1}{4}\Psi(z)\bar{D}^{2}-\frac{1}{4}\bar{\Psi}(z)D^{2})G_{V}(z,z')=-\delta^{8}(z-z'),\label{eq:GVeqn}
\end{equation}
where $\Psi\equiv W_{\Phi\Phi}$ and $D^{2}$ is the square of the
fermionic covariant derivative. Defining 
\[
\Delta\equiv(\frac{1}{4}\Psi(z)\bar{D}^{2}+\frac{1}{4}\bar{\Psi}(z)D^{2})\square^{-1}
\]
we observe that $\Delta^{2}={\cal M}^{2}\square^{-1}$ with 
\begin{equation}
{\cal M}^{2}\equiv\frac{\Psi}{4}\bar{D}^{2}\square^{-1}D^{2}\frac{\bar{\Psi}}{4}+h.c..\label{eq:M2}
\end{equation}
Thus we may write 
\begin{equation}
G_{V}=\frac{1}{\square-{\cal M}^{2}}(1+\Delta)\delta^{8}(z-z')\label{eq:GV}
\end{equation}
Note that we can write 
\begin{equation}
{\cal M}^{2}=\Psi\bar{\Psi}{\cal P}_{+}+\bar{\Psi}\Psi{\cal P}_{-}+O(D\Psi,\bar{D}\bar{\Psi})\label{eq:M2expn}
\end{equation}
where 
\[
{\cal P}_{+}=\frac{\bar{D}^{2}D^{2}}{16\square},\,{\cal P}_{-}=\frac{D^{2}\bar{D}^{2}}{16\square},
\]
 are respectively the chiral and anti-chiral projection operators.
Let us now evaluate the (anti) chiral propagators we have
\begin{eqnarray}
G_{--}(z,z') & = & D^{2}D'^{2}\frac{1}{\square-{\cal M}^{2}}(1+\Delta)\delta^{8}(z-z')\nonumber \\
 & = & D^{2}D'^{2}\frac{1}{\square-\Psi\bar{\Psi}}\left(\Psi\frac{\bar{D}^{2}}{4}\square^{-1}+O(\bar{\Psi},D\Psi,\bar{D}\bar{\Psi})\right))\delta^{8}(z-z')\label{eq:G--expn}
\end{eqnarray}
and
\begin{equation}
G_{++}(z,z')=\bar{D}^{2}\bar{D}'^{2}\frac{1}{\square-\Psi\bar{\Psi}}\left(\bar{\Psi}\frac{D^{2}}{4}\square^{-1}+O(\Psi,D\Psi,\bar{D}\bar{\Psi})\right)\delta^{8}(z-z')\label{eq:G++expn}
\end{equation}
Let us note that in the WZ model 
\begin{equation}
\Psi=m+\lambda\Phi,\label{eq:PsiWZ}
\end{equation}
and that to leading order in the derivative expansion (in terms of
$D$) the effective mass term $\Psi\bar{\Psi}$ in the denominator
of the propagators can be treated as a constant.

In the massless WZ theory it has been claimed that there is a violation
of the non-renormalization theorem stemming from a two (chiral) loop
contribution of the form 
\[
\int d^{8}z\frac{D^{2}}{\square}G(\Phi)=-\frac{1}{4}\int d^{6}z\frac{\bar{D}^{2}D^{2}}{\square}G(\Phi)=-4\int d^{6}zG(\Phi).
\]
 Such a contribution is supposed to come from the diagram in Figure
1.

In the calculation of this diagram in the literature (see \citep{Buchbinder:1994xq}
and \citep{Buchbinder:1998qv} section 4.9.5) the propagator (for
the $m=0$ theory) is taken to be \eqref{eq:G--expn} but with $\Psi\bar{\Psi}\rightarrow m^{2}=0$
in the denominator. However in actual fact even in the massless theory,
in the calculation of the effective potential, there is an effective
infra-red regulator since $\Psi\bar{\Psi}=|\lambda\Phi|^{2}\ne0$
at a generic point in field space. What is implicitly done in the
literature is to expand this denominator in powers of $\lambda$ and
keep just the coupling constant independent term. Thus the propagator
that is used is 
\[
G_{--}(z,z')=D^{2}D'^{2}\frac{1}{\square}(\Psi\frac{\bar{D}^{2}}{4}\square^{-1}+O(\bar{\Psi},D\Psi,\bar{D}\bar{\Psi}))\delta^{8}(z-z').
\]

\includegraphics[scale=0.5]{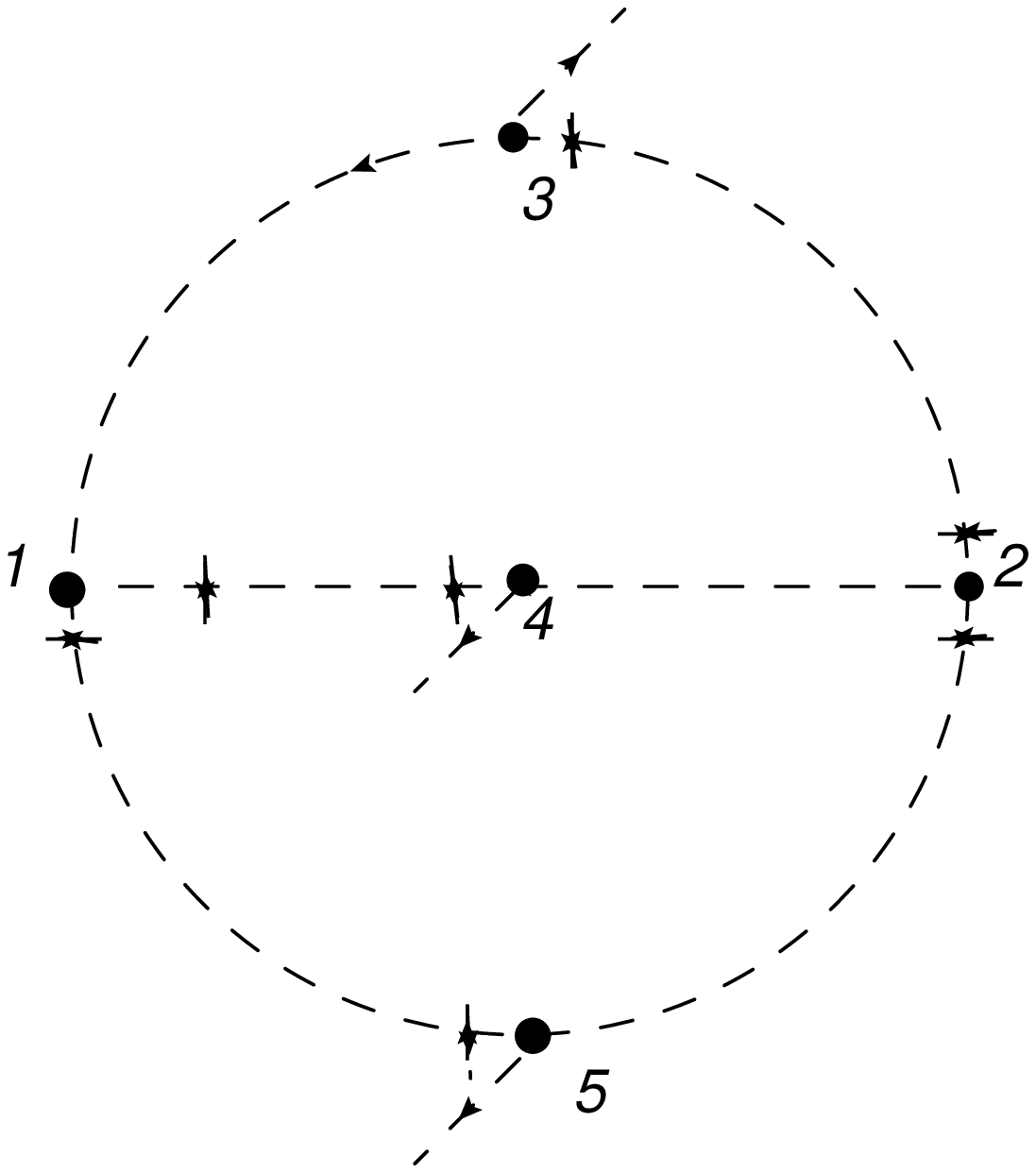}
\begin{figure}
\caption{Two loop graph contribution with insertions of $\Phi$ at points 3,4,5.
The short line segments represent insertions of $D^{2}$.}
\end{figure}

However given the fact that the effect in question is an infra red
one arising in the constant field limit ($p=0$ limit in momentum
space), the natural IR cutoff provided by $ $$\Psi\bar{\Psi}$ should
not be ignored as being $O(\lambda)$. It should instead be treated
as an IR cutoff. In other words in this infra-red situation (leading
effectively to a result which is of the form $0/0$ for the loop diagram
in the absence of a cutoff) the correct procedure should be to keep
this natural IR cutoff . The above propagator should therefore be
replaced by,
\[
G_{--}(z,z')=D^{2}D'^{2}\frac{1}{\square-|\lambda\Phi|^{2}}(\Psi\frac{\bar{D}^{2}}{4}\square^{-1}+O(\bar{\Psi},D\Psi,\bar{D}\bar{\Psi}))\delta^{8}(z-z').
\]

Given the potential for an infra-red ambiguity at $\Phi=0$ it is
crucial to keep the effective mass term in the denominator without
expanding in powers of $\lambda$, and to the extent that we are ignoring
derivative terms it is permissible to treat it as a constant effective
IR regulator $|\lambda\Phi|_{0}^{2}\equiv M^{2}$. With this propagator
the above diagram gives rise to the following contribution to the
1PI action:%
\footnote{Except for the insertion of the effective IR cutoff the calculation
is the same as that in \citep{West:1990rm,Jack:1990pd,Buchbinder:1994xq}.%
}
\begin{eqnarray}
\Gamma^{(2)} & = & -\frac{\lambda^{5}}{12}\int d^{4}x\int d^{4}\theta\int\frac{d^{4}p_{1}d^{4}p_{2}}{(2\pi)^{8}}\nonumber \\
 &  & \times\int d^{4}y_{1}d^{4}y_{2}e^{ip_{1}(x-y_{1})}e^{ip_{2}(x-y_{2})}\Phi(x,\theta)^{3}J(p_{1},p_{2})+\ldots\label{eq:Gamma(2)}
\end{eqnarray}
Note that the integration over $y_{1},y_{2}$ gives $\delta^{4}(p_{1})\delta^{4}(p_{2})$.
Here the ellipses represent derivative terms as well as non-chiral
terms, and 
\[
J(p_{1},p_{2})=\int\frac{d^{4}k_{1}d^{4}k_{2}}{(2\pi)^{8}}\frac{k_{1}^{2}p_{1}^{2}+k_{2}^{2}p_{1}^{2}-2(k_{1}.k_{2})(p_{1}.p_{2})}{\Omega(k,p)},
\]
with 
\[
\Omega=k_{1}^{2}k_{2}^{2}(k_{1}+k_{2})^{2}[(k_{1}+k_{2}-p_{1}-p_{2})^{2}+M^{2})[(k_{1}-p_{1})^{2}+M^{2}][(k_{2}-p_{2})^{2}+M^{2}].
\]
 Then we have for the integral $J\sim p^{2}/M^{2}$ so that the contribution
to the superpotential in $\Gamma^{(2)}$ vanishes (since $p_{1},p_{2}=0$
there) and thus there is no renormalization of the superpotential.
This is just the non-renormalization theorem in action in the absence
of infra-red issues. If on the other hand we had treated $M^{2}=\lambda|\Phi|^{2}$
as a perturbation and expanded in $\lambda$ then we would have got
the behavior $J\sim p^{2}/p^{2}$ resulting in a non-zero contribution
to the superpotential at the two loop level.

Thus the evaluation of this diagram with this natural IR regulator
gives zero for this purely chiral loop as would be expected for any
chiral loop in the massive WZ theory. A similar statement applies
to potential violations of the non-renormalization theorem for the
gauge coupling function (which is expected to have quantum corrections
only at one loop). These arguments will become relevant for the question
addressed in the next section on whether there is an additive contribution
to the gaugino mass in the 1PI action compared to the mass coming
from the Wilsonian action.

\section{Gaugino masses}

Let us first ignore the Weyl anomaly contribution (KL) \citep{Kaplunovsky:1993rd,Kaplunovsky:1994fg}
coming from transforming to the Einstein frame from the Jordan frame
in which off-shell supergravity is defined. To simplify the argument
let us consider a SUGRA theory with no charged scalars, no FI terms
and a single gauge group with a gauge coupling function which is just
a constant.

The full ${\cal N}=1$ supergravity with chiral scalar and gauge field
couplings was first written down by Cremmer et al \citep{Cremmer:1982en}.
In Weyl spinor formalism it is given in appendix G equation (G.2)
of Wess and Bagger (WB) \citep{Wess:1992cp} which we will use below.
Since there are many terms in this action we will not write it down
here but will refer to the relevant terms as given in this reference. 

We first observe that the action has a chiral symmetry under the ({}``$\gamma_{5}$'')
transformation $\lambda\rightarrow i\lambda$ of the gaugino field,
except for the breaking terms in lines 5 through 8, line 10 and lines
15,16, and lines 19,20. There is also the explicit (classical) gaugino
mass term in line 24. But the latter as well as lines 6,7 and 15,16,
19 and 20, are zero for constant gauge coupling functions. The contribution
of line 5 and 8 are zero under our assumption that there are no charged
chiral scalar superfields. Thus under the stated conditions the only
term which breaks the chiral symmetry of the gauginos is the dimension
5 term in line 10 i.e. 
\[
\frac{i}{4}\sqrt{g}[\psi_{m}\sigma^{ab}\sigma^{m}\bar{\lambda}+\bar{\psi}_{m}\bar{\sigma}^{ab}\bar{\sigma}^{m}\lambda](F_{ab}+\hat{F}_{ab}),
\]
 with $F_{ab},\hat{F}_{ab}$ being the gauge field strength and its
dual defined in eqn (25.17) of WB. But this term (in the absence of
non-zero background $F$) can only generate four fermi terms (and
higher powers) and cannot give mass to the gauginos. Thus in this
case a gaugino mass in the 1PI action can only arise as a consequence
of the (super) Weyl anomaly - which contains the ordinary chiral anomaly
that violates the above chiral symmetry, as well as terms related
to it by SUSY.

Let us now relax the assumption that $f$ is a constant. In this case
there will be additional terms violating the chiral symmetry $\lambda\rightarrow i\lambda$.
But any such term (in addition to the classical mass term proportional
to $F^{i}\partial_{i}f$ which vanishes when SUSY is unbroken) will
only generate terms in the 1PI effective action that are proportional
to $<\partial_{i}f>$ or higher point functions of this field. In
particular such terms will not give rise to gaugino mass terms proportional
to $m_{3/2}$ as in AMSB.

\subsection{Anomaly Effects}

The argument in the last subsection holds even in the presence of
supersymmetry breaking and has nothing to do with tuning the cosmological
constant to get flat space. Consider again the class of models where
the classical gauge coupling is constant. The gaugino sector is independent
of chiral scalars which are responsible for SUSY breaking and the
fine-tuning of the cosmological constant (by suitably adjusting a
constant term in the superpotential). In particular one can have broken
SUSY in flat space with zero gaugino mass provided there is no anomaly
in the chiral symmetry. Such an anomaly is the only way in which,
in a situation where the classical gaugino mass is zero, a quantum
one-loop mass is still generated.

The chiral symmetry (which is related to the super Weyl symmetry)
is indeed anomalous. Under a Weyl transformation characterized by
a chiral superfield transformation parameter $\tau(x,\theta)$ (which
for instance transforms $\lambda\rightarrow e^{-3\tau|_{0}}\lambda$).
The effect of this anomaly on the gauge coupling function super field
is \citep{Kaplunovsky:1993rd,Kaplunovsky:1994fg}, 
\begin{equation}
f(\Phi)\rightarrow f(\Phi)+\frac{3c}{4\pi^{2}}\tau,\label{eq:KL1}
\end{equation}
where $c=T(G)-\sum_{r}T(r)=T(G)$ since the second term, the sum over
matter representations is absent if there is no charged matter as
in our simplified case. Here $T(G)$ is the trace of a squared generator
in the adjoint representation of the gauge group. KL \citep{Kaplunovsky:1994fg}
fix the superfield $\tau$ by demanding that the transformation takes
one from the Jordan frame (which is the natural frame in which off
shell supergravity is formulated), to the Einstein frame so that $2\tau+2\bar{\tau}=-\frac{\hat{K}(\Phi,\bar{\Phi})}{3}|_{H}$.
The instruction on the LHS here is to keep only the chiral and the
anti-chiral terms in the component expansion of $K$ and is in effect
the analog of Wess-Zumino gauge fixing%
\footnote{Note that this is a superfield relation and in particular should be
used to fix the F-term of $\tau$ as well as scalar and fermionic
components. This follows from the fact that the Jordan frame superconformal
factor that needs to be removed is the superfield $e^{K/3}$.%
}. The effect of this anomaly then is to give additional terms to both
the gauge coupling and and the gaugino mass%
\footnote{We ignore an additional (NSVZ) anomaly term coming from redefining
the gauge kinetic term to get canonical normalization for it.%
},
\begin{eqnarray}
\frac{1}{g_{phys}^{2}} & = & \Re f-\frac{3T(G)}{16\pi^{2}}K|_{0},\label{eq:KL2}\\
\frac{2M}{g_{{\rm phys}}^{2}} & = & \frac{1}{2}F^{i}\partial_{i}f-\frac{3T(G)}{16\pi^{2}}F^{i}K_{i}|_{0}.\label{eq:KL3}
\end{eqnarray}
In SQCD coupled to supergravity with neutral chiral scalars breaking
SUSY and a classical gauge coupling constant which is field independent,
the first term on the RHS of the gaugino mass equation \eqref{eq:KL3}
will be zero, but there will nevertheless be a mass term that is generated
by the Kähler-Weyl anomaly. This term however also vanishes if supersymmetry
is unbroken (i.e. $F=0$) as needs to be the case in AdS supersymmetry.
It also satisfies the criterion that in the Wilsonian effective action
UV effects should not break the structure of the supergravity action
given in appendix G of \citep{Wess:1992cp}. In other words unless
one is claiming that there is an anomaly in local supersymmetry the
general structure of the local supergravity action should be preserved%
\footnote{For a detailed discussion of these issues see \citep{deAlwis:2012gr}.%
}, with the appropriate perturbative corrections to the expressions
for the Kähler potential, the superpotential and the gauge coupling
function. 

This is in contrast to the claims in the AMSB literature \citep{Randall:1998uk,Giudice:1998xp},
where an additional term $\frac{3T(G)}{16\pi^{2}}e^{K/2}W|_{0}$ is
said to be needed on the RHS of \eqref{eq:KL3}. Note that this term
is independent of the factor $\partial_{i}f$ and therefore has nothing
to do with possible corrections to gaugino mass in the 1PI action
coming from non-mass terms in the Wilsonian action proportional to
this factor.

There is also the possibility in principle that IR effects in the
1PI action can violate arguments based on the Wilsonian action as
in the above discussion. However as we've argued in the previous section
such violations are spurious. Thus our conclusion is that if the gaugino
mass in the Wilsonian effective action (after including the Weyl anomaly
contribution) is zero (as would be the case if SUSY is unbroken),
then the physical gaugino will have zero mass.

\subsection{Non-perturbative effects}

In a recent paper \citep{Dine:2013nka} it has been argued that there
is a non-perturbative effect coming from gaugino condensation which
requires that above the scale of the condensing gauge group (say $\Lambda_{c}$)
one should have added a {}``anomaly mediated'' gaugino mass counter
term in the effective theory. If true this claim would appear to violate
the above argument. So let us examine it.

The argument proceeds from the observation that when there is a non-perturbatively
generated superpotential $W_{np}$, well below the scale $\Lambda$,
we should be able to identify the term 
\begin{equation}
-3W_{np}W_{0}^{*}+h.c.\label{eq:W0Wnp}
\end{equation}
in the potential with $W_{0}$ being a constant in the superpotential.
It is then claimed that the existence of this term requires adding
the {}``anomaly-mediated'' gaugino mass counter term to the high-energy
theory,
\[
{\cal L}_{\lambda\lambda}=\frac{1}{2}\frac{3T(G)}{16\pi^{2}}e^{K/2}W|_{0}\lambda\lambda+h.c.
\]

Firstly, below the scale of the condensing gauge group (but above
the SUSY breaking soft mass scale) obviously there are no gauginos
anymore that pertain to the gauge group under consideration. Since
the degrees of freedom of this gauge group have been integrated out,
they are irrelevant to the low energy phenomenology. This is in contrast
to the claim of AMSB which posits a contribution (proportional to
the gravitino mass) to the mass of a gaugino which survives in the
low energy theory. 

Secondly the argument is based on reasoning which is not valid for
the off-shell formulation of supergravity (which is what one needs
at the quantum level). In the off-shell formulation (i.e. before integrating
out auxiliary fields), the theory is linear in the superpotential.
Indeed as long as supersymmetry is not broken explicitly, the superpotential
dependent part of the superspace action (with chiral superspace measure
${\cal E}$) must take the form%
\footnote{It is convenient for the subsequent discussion to keep explicitly
the chiral scalar compensator in the action as in \citep{Kaplunovsky:1994fg}.%
} 
\begin{equation}
\int d^{6}z{\cal E}C^{3}W(\Phi).\label{eq:SuperW}
\end{equation}
The total superpotential is now a sum of the classical and non-perturbative
terms,
\begin{equation}
W=W_{cl}(\Phi)+W_{NP}=W_{cl}(\Phi)+Ae^{-3\frac{8\pi^{2}}{b_{c}}f_{c}(\Phi)}.\label{eq:Wtotal}
\end{equation}
For simplicity we've assumed in the above that there are no matter
fields charged under the condensing gauge group $G_{c}$ (with gauge
coupling function $f_{c}$). Its 1 loop beta function coefficient
is $b_{c}=3T(G_{c})$. The term \eqref{eq:W0Wnp} then arises in the
usual fashion from the term $-3|W|^{2}$ that comes in the potential
once one eliminates the auxiliary fields to get the on-shell supersymmetric
action.

The manner in which the second term of \eqref{eq:Wtotal} arises from
a condensing gauge group in the supergravity context was discussed
first in \citep{Burgess:1995aa} and elaborated on in \citep{deAlwis:2012bm}.
Let us recapitulate the argument as presented in the latter reference.
Above the confining scale $\Lambda_{c}$ of $G_{c}$ the action has
an explicit superspace gauge field kinetic term $\left(\int d^{6}z{\cal E}\frac{1}{4}f_{c}(\Phi){\cal W}^{c}{\cal W}^{c}+h.c.\right)$.
Well below the scale $\Lambda_{c}$ the $G_{c}$ degrees of freedom
need to be integrated out. This gives an effective action $\Gamma$
defined schematically by 
\begin{equation}
e^{-\Gamma(\Phi,\bar{\Phi},C,\bar{C})}=\int d(gauge)\exp\left\{ -\frac{1}{4}\int[f_{c}(\Phi)-\frac{b_{c}}{8\pi^{2}}\ln C]{\cal W}^{c}{\cal W}^{c}+h.c.\right\} ,\label{GammaEffective}
\end{equation}
where we included the KL anomaly contribution\citep{Kaplunovsky:1994fg}
corresponding to\eqref{eq:KL1} in the gauge coupling function. Since
SUSY should not be broken by this procedure, we expect $\Gamma$ %
\footnote{The crucial assumption here is the quasi-locality of this contribution
to $\Gamma$ which enables us to define its derivative expansion and
then focus on its two derivative action which should be be of the
standard supergravity form.%
} to have the general form of a superspace action and in particular
should develop a superpotential. Given the general argument that any
superpotential should come with a factor $C^{3}$, we see that the
corresponding term in $\Gamma$ will be (a superspace integral of)
\begin{equation}
C^{3}W_{NP}=C^{3}A_{a}\exp\left(-3\frac{8\pi^{2}}{b_{a}}f_{a}(\Phi)\right).\label{eq:WNP}
\end{equation}
The total superpotential below $\Lambda_{c}$ is then given by \eqref{eq:Wtotal}.
Thus there is absolutely no need to add any kind of counter term proportional
to $W_{cl}$ to the gaugino mass in the UV theory above the scale
$\Lambda_{c}$. Indeed the addition of such a term would be a violation
of the principle that the Wilsonian action at the two derivative level
must preserve the general structure of a SUGRA action given in appendix
G of WB \citep{Wess:1992cp}. It should also be stressed that once
the auxiliary fields are solved for in terms of the propagating fields,
the Einstein-Kähler frame Lagrangian will be exactly as given in equation
G2 of WB \citep{Wess:1992cp} with the classical superpotential being
replaced everywhere by the sum of the classical and non-perturbative
terms \eqref{eq:Wtotal}. In particular this would mean of course
an additional contribution to the gravitino mass so that line 21 of
equation G2 of WB will become

\begin{equation}
-ee^{K/2}\{(W_{cl}+W_{NP})\bar{\psi}_{a}\bar{\sigma}^{ab}\bar{\psi}_{a}+h.c.\}.\label{eq:gravitinomass}
\end{equation}

\subsection{Linearized SUGRA}

In another recent paper \citep{DiPietro:2014moa} the authors have
given what on the face of it appears to be yet another argument for
an additive term in the expression for the gaugino mass. However this
term appears only because the linearized formalism in which the authors
work ignores a term coming from the chiral density ${\cal E}$ (which
in the authors' formalism has been set to unity)%
\footnote{In the traditional argument for AMSB given in \citep{Randall:1998uk,Giudice:1998xp}
for instance a similar problem arises because of the linearized formalism
in which the authors work. For detailed discussion see \citep{deAlwis:2012gr}.
For a discussion of other issues that may be missed in the linearized
formalism as compared to the full non-linear supergravity see \citep{deAlwis:2012aa}.%
}. In other words the correct superspace term corresponding to their
equations (28) and (30) should have the form, in the notation of WB
\citep{Wess:1992cp} (for a constant gauge coupling function such
as the one implicit in these two equations), $c\int d^{2}\theta{\cal E}{\cal W}{\cal W}$.
The authors only calculate the F-term of ${\cal W}^{2}$ but ignore
(since for them ${\cal E}=1$) the F-term of the chiral density which
exactly cancels the term that they calculate. Indeed this is why in
equation G2 of WB, when the gauge coupling function is a constant,
there is no gaugino mass term. The penultimate line in this equation
$ $is zero, whether or not one is in flat space (with ${\cal E}=1$).
the point is the action must first be calculated in a generally covariant
and supersymmetric fashion, before one specializes to flat space thus
breaking general covariance. The authors are therefore in manifest
contradiction with the Cremmer et al Lagrangian \citep{Cremmer:1982en}
given in equation G2 of \citep{Wess:1992cp}.

As pointed out in the earlier works \citep{deAlwis:2012gr}\citep{deAlwis:2008aq}
by the author, the only way an additive gaugino mass contribution
(i.e. one that does not vanish when the the F-term of chiral scalars
is zero), can arise is if there is an anomaly in local supersymmetry.
Indeed preserving general covariance as well as supersymmetry is crucial
for getting correct results. If there is no anomaly in either general
covariance or supersymmetry, then the Wilsonian effective action (including
the effects of integrating out high scales) should respect the structure
of the classical theory. This means that the action is still given
in terms of a Kähler potential ($K)$ a superpotential $W$ and a
gauge coupling function $f$ with appropriate loop and non-perturbative
corrections. What we have argued in this section is that the physical
masses (position of poles in propagators) will be perturbatively related
to those in the Wilsonian action. In particular there cannot be an
additive contribution to the gaugino mass as compared to the corresponding
mass term in the Wilsonian action.

\section{Phenomenological issues and Conclusions}

In phenomenology based on high scale supersymmetry breaking in the
hidden sector of a SUGRA model, the (classical) masses and couplings
calculated at that scale are used as initial conditions for RG evolution
down to the TeV scale. In effect these running masses and couplings
are computed as Wilsonian parameters - and are then compared to data
at the TeV scale for scalar as well as gaugino masses. But these are
obviously not the actual parameters in the 1PI action - in particular
the physical mass is the position of the pole (or its real part for
a unstable particle) in the two point function. However the difference
between the Wilsonian mass and the 1PI mass is a higher order effect
and is smaller than the effect of running. The point of running down
to the TeV scale is of course the expectation that the physical (pole)
mass is near that scale and that the (small) difference can be computed
in perturbation theory.

In standard renormalization theory mass (and coupling) parameters
are input data from experiment. For instance the physical electron
mass (defined as the zero of the inverse propagator) is an experimental
input which goes into the renormalized Lagrangian. Similarly the coupling
is defined at some scale in terms of a cross section at that scale
- in QED for example this would be fixed by the Thompson cross section.
To compute the cross section at some high scale, one would use the
beta function equations to evolve the coupling and mass parameter
to the appropriate scale using these physical inputs as initial conditions.
These new parameters are then used in a perturbative calculation of
a cross section at that high scale.

In making predictions for SUSY soft masses and couplings we must necessarily
follow the inverse procedure. Firstly of course we need to assume
that low energy (TeV scale) supersymmetry exists! This means that
we expect experiments to yield partners to the standard model particles
at the this scale. But experiments give us the value of physical i.e.
pole masses for the lightest superparticle (LSP), the stop, the gluino,
etc. The theory of supersymmetry breaking however has a natural scale
which is typically the scale at which we can construct a SUGRA coming
say from string theory. In most scenarios this is around the string/GUT
scale. The theory then gives a value for the masses etc. which are
then used as initial conditions for evolving down to the TeV scale.
Assuming that SUSY is relevant for solving the hierarchy problem,
one evolves the RG equations down to a scale of say a TeV. Thus one
expects the (hopefully) measured masses - the pole masses in the S-matrix
- to be given by the running masses at this scale, up to small perturbative
corrections. To put it another way the Wilsonian masses should be
the leading approximation to the actual physical masses in the quantum
effective action, with any differences being perturbatively small. 

We have addressed in this paper two issues that are relevant to this
procedure. The first is that in computing the 1PI action one should
work at a generic point in field space. The (non-zero) field then
acts as an effective IR regulator so that terms which in the absence
of a regulator violate the non-renormalization theorems, are in fact
absent. Thus we expect all the constraints of supersymmetry that are
present in the Wilsonian effective action to be present in the local
terms (such as the mass terms) of the 1PI action. Secondly we've argued
that the physical (pole) mass of fermions (in particular gauginos)
cannot acquire an additive contribution as compared to the mass term
in the Wilsonian effective action. Finally we addressed some claims
in two recent papers which appear to get a contribution to the gaugino
mass that does not fit the structure of the generally covariant and
supersymmetric effective action.

Let us  reemphasize the main message of this note. In deriving a quantum
effective action for locally supersymmetric theories, one should keep
natural infrared cutoffs and not break general covariance by working
in some fixed metric background. Failure to do so can give rise to
spurious effects, i.e. ones which are absent in the generally covariant
and supersymmetric action such as the so-called AMSB term in the gaugino
mass%
\footnote{It is a curious fact that in string models with no-scale like structure
for the Kaehler potential there is a cancellation between this AMSB
term and the KL term - i.e. the second term in \eqref{eq:KL3}. Since
practically all of string phenomenology is based on such Kaehler potentials,
and given that these are the only UV completions that we know of for
theories for the MSSM soft masses, this would mean that if one imposed
the criterion that a sensible theory of SUSY breaking should have
an UV completion, then as far as we know the adherents of the AMSB
term must admit that the whole Weyl anomaly issue is spurious. On
the other hand it has long been the contention of this author \citep{deAlwis:2008aq,deAlwis:2012gr}
that since only the KL term is present so that there is no cancellation,
the Weyl anomaly is indeed very relevant for a discussion of SUSY
phenomenology in string theory.%
}.

\section{Acknowledgements}

I wish to acknowledge discussions with Oliver DeWolfe and Michael
Dine and Jan Louis. This research is partially supported by the United
States Department of Energy under grant DE-FG02-91-ER-40672. 

\bibliographystyle{apsrev}
\bibliography{myrefs}

\end{document}